\documentclass[usenatbib]{mn2e}

\usepackage[dvips]{graphicx}
\usepackage{amssymb}
\usepackage{txfonts}

\newcommand{\msun}{{\rm M}_{\sun}}

\newcommand{\g}{$\gamma$}

\newbox\grsign \setbox\grsign=\hbox{$>$} \newdimen\grdimen \grdimen=\ht\grsign
\newbox\simlessbox \newbox\simgreatbox \newbox\simpropbox
\setbox\simgreatbox=\hbox{\raise.5ex\hbox{$>$}\llap
   {\lower.5ex\hbox{$\sim$}}}\ht1=\grdimen\dp1=0pt
\setbox\simlessbox=\hbox{\raise.5ex\hbox{$<$}\llap
   {\lower.5ex\hbox{$\sim$}}}\ht2=\grdimen\dp2=0pt
\setbox\simpropbox=\hbox{\raise.5ex\hbox{$\propto$}\llap
   {\lower.5ex\hbox{$\sim$}}}\ht2=\grdimen\dp2=0pt
\def\ga{\mathrel{\copy\simgreatbox}}
\def\la{\mathrel{\copy\simlessbox}}

\title[Hadronic models of blazars]{Hadronic models of blazars require a change of the accretion paradigm}

\author[A. A. Zdziarski \& M. B{\"o}ttcher]
{Andrzej A. Zdziarski$^1$ and Markus B{\"o}ttcher$^{2,3}$\\
$^1$Centrum Astronomiczne im.\ M. Kopernika, Bartycka 18, PL-00-716 Warszawa, Poland\\
$^2$Centre for Space Research, North-West University, Potchefstroom 2520, South Africa\\
$^3$Astrophysical Institute, Department of Physics and Astronomy, Ohio University, Athens, OH 45701, USA
}

\date{Accepted 2015 March 5.  Received 2015 March 3; in original form 2015 January 24}

\pagerange{\pageref{firstpage}--\pageref{lastpage}}
\pubyear{2015}

\begin{document}

\maketitle

\label{firstpage}

\begin{abstract}
We study hadronic models of broad-band emission of jets in radio-loud active galactic nuclei, and their implications for the accretion in those sources. We show that the models that account for broad-band spectra of blazars emitting in the GeV range in the sample of B{\"o}ttcher et al.\ have highly super-Eddington jet powers. Furthermore, the ratio of the jet power to the radiative luminosity of the accretion disc is $\sim 3000$ on average and can be as high as $\sim 10^5$. We then show that the measurements of the radio core shift for the sample imply low magnetic fluxes threading the black hole, which rules out the Blandford-Znajek mechanism to produce powerful jets. These results require that the accretion rate necessary to power the modelled jets is extremely high, and the average radiative accretion efficiency is $\sim 4 \times 10^{-5}$. Thus, if the hadronic model is correct, the currently prevailing picture of accretion in AGNs needs to be significantly revised. Also, the obtained accretion mode cannot be dominant during the lifetimes of the sources, as the modelled very high accretion rates would result in too rapid growth of the central supermassive black holes. Finally, the extreme jet powers in the hadronic model are in conflict with the estimates of the jet power by other methods. 
\end{abstract}

\section{Introduction}
\label{intro}

The most commonly considered models of broad-band emission of radio-loud jets in AGNs utilize emission of relativistic electrons and positrons via synchrotron and Compton processes. These leptonic models have been highly successful in reproducing the spectra and time variability of blazars and radio galaxies, though there are some phenomena, such as their extremely short variability time scales, in some cases down to a few minutes 
\citep{Aharonian07,Albert07}, that are not readily explained by them.

The alternative, and quite popular, model is based on hadronic (or lepto-hadronic) processes (e.g., \citealt{mb92,mp01,aharonian00}). It utilizes emission of extremely relativistic ions, mainly due to proton-synchrotron and photo-pion production processes, to explain the high-energy emission from jet-dominated AGNs. The latter process gives rise to cascades of electrons and positrons, also emitting via the synchrotron and Compton processes. The peak of the high-energy emission of the broadband spectral energy distribution of blazars, at $\sim 0.1$--10 GeV, is then due to these processes, mostly proton synchrotron. 

We stress that the name 'hadronic' refers only to the emission processes, and the 'leptonic' jets in AGN models also contain ions, though with relatively low energies, preventing them from radiating efficiently. Still, those hadrons in the leptonic models usually dominate the kinetic luminosity of jets (e.g., \citealt{ghisellini14}). 

\citet{sikora09} and \citet{sikora11} have shown that the hadronic processes are very inefficient. This implies that, if the total jet power is approximately limited by the Eddington luminosity, the hadronic model can be ruled out in many cases. On the other hand, \citet{boettcher13} (hereafter B13) have successfully applied hadronic models to a sample of radio-loud AGNs circumventing this constraint by allowing the jet power to be highly super-Eddington. Here we discuss consequences of this supposition.

\section{Analysis of the sample of B13}
\label{sample}

We study here the sample of 12 blazars of B13. Their broad-band spectra were fitted by them by leptonic and hadronic models, and we consider here the jet powers obtained by B13 for the latter. B13 provide the redshifts, $z$, the apparent superluminal velocity, $\beta_{\rm app}$, and the accretion luminosity, $L_{\rm acc}$, see Table 1. Six of those objects were also present in the sample of \citet{z14} (hereafter ZCST14), for which those authors provide estimates of the black-hole mass, $M$, $\beta_{\rm app}$, and $L_{\rm acc}$ based on literature. The values of $\beta_{\rm app}$ agree well between ZCST14 and B13, while the values of $L_{\rm acc}$ agree to within a factor of $\sim$2, which is a satisfactory agreement given the necessary approximate character of those estimates. However, the exception is W Comae, for which the $L_{\rm acc}$ estimate used by B13 is 40 times higher than that given in ZCST14, which is the upper limit of \citet{ghisellini10}, hereafter G10. B13 used the estimate of \citet{xie08}, who refer in turn to two other papers, which do not seem to give that information. Thus, we use here the value of G10. B13 quote an estimate of $L_{\rm acc}$ of \citet{xie08} for OJ 287, which is 7 times higher than the upper limit of G10; here we use the latter. For two objects (S5 0716+714 and 3C 66A), B13 give no estimates of $L_{\rm acc}$, and we use the upper limits given by G10. In the case of 3C 66A, there is also an uncertainty of its redshift, and B13 used an estimated redshift of $z=0.3$ based on EBL-corrected SED modelling \citep{Abdo11}, which is also marginally consistent with the spectroscopically determined lower limit on the redshift by \citet{Furniss13}. Since we use the estimates of the jet power from B13, we adopt their redshift value and rescale the upper limit of $L_{\rm acc}$ of G10, who used $z=0.444$, by the ratio of $D_L^2$, where $D_L$ is the luminosity distance. We assume the same cosmological parameters as B13, $\Omega_\Lambda=0.7$, $\Omega_{\rm m}=0.3$ and $H_0=70$ km/(s \,Mpc). 

For the objects not in the sample of ZCST14, we find mass estimates from the literature, in particular from G10. In some cases, there is a considerable uncertainty. For OJ 287, we use $6.2\times 10^8\, \msun$ from \citet*{wang04}. However, if the system is indeed a binary black-hole system, the mass of the less and more massive component is estimated as $1.4\times 10^{8}\, \msun$ and $1.8\times 10^{10}\, \msun$, respectively \citep*{valtonen12}. Accretion would be dominated by the more massive black hole. The adopted values of $L_{\rm acc}$ and $M$ for all sources studied here are given in Table 1. 

\begin{figure}
\centerline{\includegraphics[width=5.5cm]{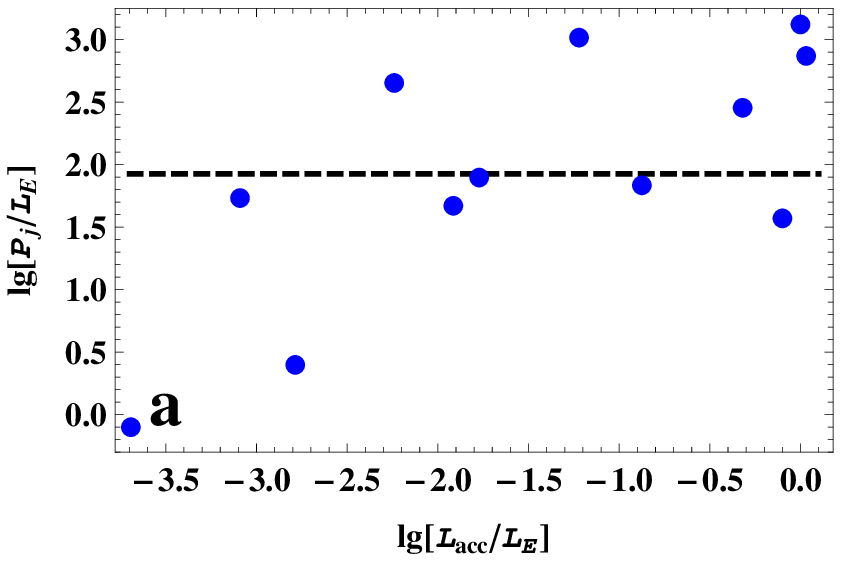}}
\centerline{\includegraphics[width=5.5cm]{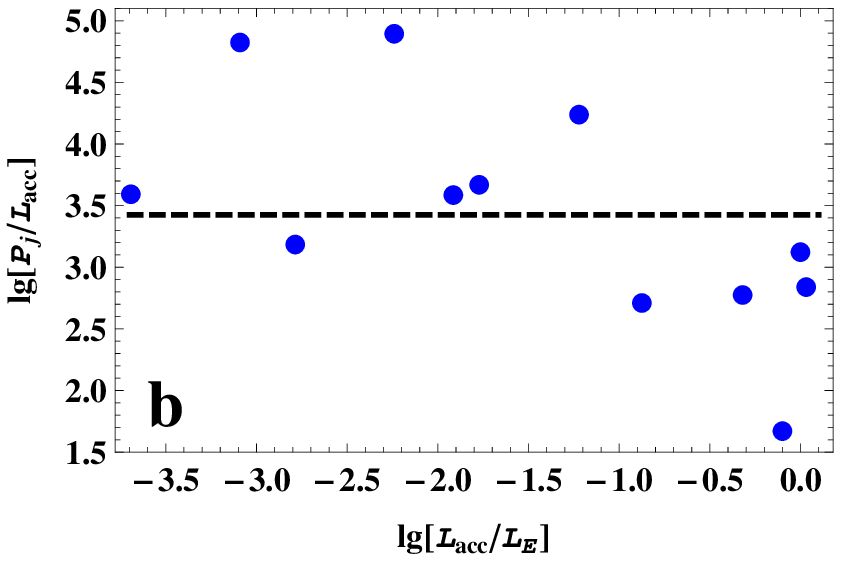}}
\caption{The ratio of the total jet power to (a) the Eddington luminosity and (b) the accretion luminosity as functions of the Eddington ratio. The dashed lines show the geometric averages.
}
\label{power}
\end{figure}

\setlength{\tabcolsep}{5pt}
\begin{table*}
\begin{center}
\caption{The main parameters of the sample. FSRQ - flat spectrum radio quasar, LBL, IBL and HBL refer to blazars which synchrotron spectrum has the peak at low, intermediate and high frequencies, respectively; $\nu$ and $\nu F_\nu$ give the estimated position of the peak of the GeV spectrum approximated as a power law, and $P_{\rm j}$ is the total jet+counterjet power in the hadronic model of B13. See B13 for the remaining parameters of the sources.}
\begin{tabular}{lccccccccccc}
\hline
Object & type &$z$ &$L_{\rm acc}$ & $M$ &$L_{\rm acc}/L_{\rm E}$& $\Delta\theta$ &$\nu$ &$\nu F_\nu$ & $\gamma_{\rm min}$ & $\gamma_{\rm max}$ & $P_{\rm j}$\\ 
&&&erg\,s$^{-1}$ &$\msun$ && mas & $10^{22}\,{\rm Hz}$ & $10^{-10}{\rm erg\,cm}^{-2}{\rm s}^{-1}$ && $10^9$ & $10^{48}\,{\rm erg\,s}^{-1}$ \\
\hline
0219+428 (3C 66A) &IBL	&0.3 &$4.80\!\times\! 10^{43}$ &$4\!\times\! 10^{8}$  &$8.1\!\times\! 10^{-4}$ &0.073&10 &1 &1 &1.2 &3.2\\
0235+164 (AO)	 &LBL	&0.94 &$3.40\!\times\! 10^{44}$ &$4\!\times\! 10^{8}$  &0.0057         &0.006&30 &7  &1 &4.3 &27\\
0420-014 (PKS)	 &FSRQ	&0.914&$3.02\!\times\! 10^{46}$ &$2.57\!\times\!10^{8}$ &0.97           &0.256&5 &1 &$10^3$ &0.43 &1.4\\
0528+134 (PKS)	 &FSRQ	&2.07 &$1.70\!\times\! 10^{47}$ &$1.07\!\times\!10^{9}$ &1.07          &0.150&5 &2  &1 &1.1 &117\\
0716+714 (S5)	 &LBL	&0.31 &$1.80\!\times\! 10^{44}$ &$1\!\times\! 10^{8}$  &0.012          &0.125&70 &1 &1 &2.7 &0.69\\
0851+202 (OJ 287) &LBL	&0.306&$1.50\!\times\! 10^{44}$ &$6.2\!\times\! 10^{8}$ &0.0016          &0.051&5 &0.5&1 &1.0 &0.23\\
1219+285 (W Comae)&IBL &0.102&$1.50\!\times\! 10^{43}$ &$5\!\times\! 10^{8}$  &$2.0\!\times\! 10^{-4}$ &0.047&80 &0.5&1 &1.9 &0.059\\
1226-023 (3C 273) &FSRQ	&0.158&$1.30\!\times\! 10^{47}$ &$6.59\!\times\!10^{9}$ &0.13           &0.017&2 &10 &$10^3$ &0.43 &67\\
1253-055 (3C 279) &FSRQ	&0.536&$2.00\!\times\! 10^{45}$ &$8\!\times\! 10^{8}$  &0.017          &0.051&10 &1 &$10^3$ &0.63 &9.3\\
1510-089 (PKS)	 &FSRQ	&0.36 &$1.12\!\times\! 10^{46}$ &$1.58\!\times\!10^{8}$ &0.48           &0.151&8 &5 &$10^3$ &1,1 &6.7\\
2200+420 (BL Lac) &LBL	&0.069&$1.51\!\times\! 10^{45}$ &$1.70\!\times\!10^{8}$ &0.060          &0.052&10 &0.3&1 &1.9 &26\\
2251+058 (3C454.3)&FSRQ	&0.859&$7.24\!\times\! 10^{46}$ &$4.90\!\times\!10^{8}$ &1.00           &0.159&10 &10 &$10^3$ & 1.1 &96\\
\hline
\end{tabular}
\end{center}
\label{pars}
\end{table*}

B13 give the jet power\footnote{The entry for $L_{\rm p}$ (the power in protons) of OJ 287 in B13 is a typo, it should be 0.083 rather than 8.3.} estimated using the energy content and for one jet only, see their equations (3--5). However, the jet power, $P_{\rm j}$, is the enthalpy flux (e.g., \citealt{levinson06}), and the counterjet should be included in the energy budget. Thus, we multiply their values by 8/3, assuming the protons are relativistic (and thus their equation of state can be described by an adiabatic index of 4/3), see Table 1. We neglect a possible contribution of cold protons, and consider only those in the power-law distribution assumed to give rise to the observed spectrum, following the fits of B13. We then compare $P_{\rm j}$ to both $L_{\rm E}$ (calculated for an H fraction of $X=0.7$) and $L_{\rm acc}$ in Fig.\ \ref{power}. We find $\langle P_{\rm j}/L_{\rm E}\rangle\simeq 85$, $\langle P_{\rm j}/L_{\rm acc}\rangle\simeq 2660$, with the standard deviations corresponding to factors of $\simeq$10 and $\simeq$8.4, respectively. (Hereafter the symbol $\langle . \rangle$ denotes a geometric average, i.e., average of logarithms, and the given standard deviation corresponds to a multiplicative factor.) Except for W Comae and OJ 287, the distribution of $P_{\rm j}/L_{\rm E}$ is relatively uniform. As a consequence, there is an anti-correlation between $P_{\rm j}/L_{\rm acc}$ and $L_{\rm acc}/L_{\rm E}$, but all the obtained values are still very high, spanning $P_{\rm j}/L_{\rm acc}\sim 50$--$10^5$. Furthermore, some of our values of $L_{\rm acc}$ are upper limits, and the corresponding true values of $P_{\rm j}/L_{\rm acc}$ can be even higher.

In general, we can write $P_{\rm j}=\epsilon_{\rm j} \dot M c^2$, where $\dot M$ is the accretion rate and $\epsilon_{\rm j}\la 1.5$, the ratio of the jet power to the available accretion power, which we will call here the jet formation efficiency. The approximate maximum corresponds \citep*{tnm11,mtb12} to efficient extraction of the black-hole spin energy \citep{bz77}, in which case the jet power may exceed $\dot M c^2$. However, attaining such a high jet formation efficiency requires the accretion flow to possess a strong poloidal magnetic field \citep{tnm11,mtb12}, namely to form a magnetically arrested disc (MAD, \citealt*{nia03}; see also \citealt{bkr76}). The required energy density corresponds then to the magnetic flux threading a rapidly spinning black hole of $\Phi_{\rm BH}=\phi_{\rm BH}(\dot Mc)^{1/2} r_{\rm g}$, with $\phi_{\rm BH}\simeq 50$, which corresponds to the magnetic field strength at the horizon of $B^2/8\upi\sim 100 \dot m m_{\rm p} c^2/(\sigma_{\rm T} r_{\rm g})$, where $\dot m=\dot M c^2/L_{\rm E}$, $r_{\rm g}=G M/c^2$ is the gravitational radius, $m_{\rm p}$ is the proton mass, and $\sigma_{\rm T}$ is the Thomson cross section. Since the magnetic flux is conserved in the jet, it can be measured at large scales, e.g., through the radio-core shift effect \citep{lobanov98,hirotani05}. Values of $\phi_{\rm BH}\sim 50$ were found for a large sample of blazars and radio galaxies by ZCST14 assuming $\dot M c^2=L_{\rm acc}/\epsilon_{\rm r}$ for their adopted accretion radiative efficiency of $\epsilon_{\rm r}=0.4$. However, given that $P_{\rm j}\gg L_{\rm acc}$ in the hadronic model of B13, we have to estimate $\dot M$ instead from the jet power, strongly dominating the energy budget, i.e., $\dot M c^2\simeq P_{\rm j}/\epsilon_{\rm j}\gg L_{\rm acc}/\epsilon_{\rm r}$.

\begin{figure}
\centerline{\includegraphics[width=5.5cm]{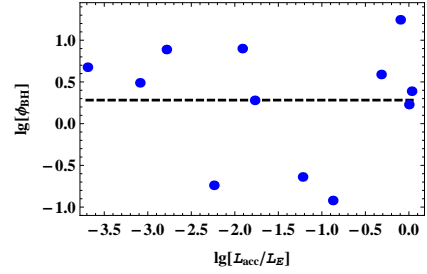}}
\caption{The dimensionless magnetic flux $\phi_{\rm BH}$ as a function of the Eddington ratio. The dashed line show the geometric average of $\phi_{\rm BH}\simeq 1.9$. The model of efficient jet formation via black-hole spin energy extraction \citep{bz77} predict $\phi_{\rm BH}\ga 50$ \citep{tnm11,mtb12}, which is much higher than the obtained values.
}
\label{phi}
\end{figure}

To calculate the values of $\phi_{\rm BH}$ implied by the hadronic model, we use the radio core shift, $\Delta\theta$, measurements between 8 and 15 GHz for the sources in our sample from \citet{pushkarev12}, which we give in Table 1. We apply a correction of \citet{zspt15} to the power of the $(1+z)$ factor in the formula for the magnetic field strength at 1 pc, $B_1$, and then use the formula for the jet magnetic flux, $\Phi_{\rm j}$, of equation (5) of ZCST14. From $\Phi_{\rm j}=\Phi_{\rm BH}$ we obtain $\phi_{\rm BH}$, which we plot in Fig.\ \ref{phi} for $\epsilon_{\rm j}=1$ and the dimensionless black-hole angular momentum of 1. We find $\langle\phi_{\rm BH}\rangle\simeq 1.9$ with the standard deviation of a (multiplicative) factor of $\simeq 5$. These result can be compared to another formula for $\Phi_{\rm j}$ of \citet{zspt15}, which takes into account effects due to the small jet opening angles in the radio-emitting region (and the consequent low value of the magnetization parameter), and due to transverse averaging of the magnetic field. That formula yields values lower by approximately $\sqrt{2}$ from those above. Thus, our results appear to give robust approximate estimates of the magnetic flux (under the assumption of the hadronic jet model). Given this and that {\it all\/} objects in the sample have $\phi_{\rm BH}\ll 50$, we can rule out the efficient version of the model of \citet{bz77} for the sample. (We note that this model is likely to work for the studied sources if the leptonic model is adopted, as in ZCST14.)

In other jet formation models (e.g., \citealt{bp82,cb14}), the jet formation efficiency has to be $<1$, and it is usually $\epsilon_{\rm j}\ll 1$. Thus, the accretion process is even more highly super-Eddington. For such accretion, \citet{cb14} make a rough estimate of $\epsilon_{\rm j}=0.1$, which we also adopt. Then, $\langle \dot M c^2/L_{\rm E}\rangle\sim 10^3$, i.e., the accretion is indeed highly supercritical. The radiative efficiency of the accretion disc in this case is very small, $\langle L_{\rm acc}/\dot M c^2\rangle \simeq 4\times 10^{-5}$. 

After the calculations presented in this paper had been completed, \citet{cerruti15} presented an analysis of 5 high-frequency peaked BL Lac objects (HBLs), whose \g-ray peak is located around $\sim$1 TeV rather than at $\sim$1 GeV as typical for LBLs and IBLs considered in B13 and in this work. \citet{cerruti15} found that it is possible to find hadronic-model fits with sub-Eddington jet powers for those particular objects. A detailed analysis of those objects is outside the scope of this Letter, but we briefly address a few issues. \citet{cerruti15} took into account secondary products of hadronic processes in more detail than B13. However, they assumed steady state primary proton and electron distributions as fixed power laws and followed the evolution of particle spectra, using the respective kinetic equations, only for the secondary particles, while B13 used the electron and proton kinetic equations for all particles, which is the self-consistent approach. It is not clear to us how this might affect the jet powers. Also, the jet powers given in \cite{cerruti15} are for one jet only, neglect the pressure contribution, and assume the jet Lorentz factor, $\Gamma$, to be one-half of the Doppler factor, $\delta$, while $\Gamma=\delta$ for the inclination angle $i=1/\Gamma$, as favoured on statistical grounds. Thus, for comparison with the results here, their jet powers need to be multiplied by a factor of 32/3. Still, we do not exclude that a sub-Eddington jet powers can be obtained for some objects, as it is, e.g., the case for W Comae in our sample. Generally, due to the substantially lower total luminosities and at least equally large (if not larger) black-hole masses of HBLs compared to low-frequency peaked blazars (especially FSRQs), it seems less problematic to produce hadronic model fits with sub-Eddington jet powers for HBLs. 

\section{The minimum jet power in the proton-synchrotron model}
\label{minimum}

The above results are for the model fits of B13 which, due to the large number of parameters, may generally not be the only possible hadronic model representation of the chosen SEDs. We may therefore ask whether another set of fits of the hadronic model could have lower total power. We can answer this question by finding the minimum possible power for a given proton synchrotron flux, found to dominate the overall model spectra of the objects in the sample, see fig.\ 9 of B13. We use the method of \citet{zdz14} (hereafter Z14), who calculated the minimum jet power for a given electron synchrotron flux. We can adapt those calculations to the proton synchrotron case by replacing the electron mass, $m_{\rm e}$, by $m_{\rm p}$ in the relevant formulae of Z14 (including physical constants). Equivalently, $\sigma_{\rm T}$, the critical magnetic field, $B_{\rm cr}$, the dimensionless photon energy in the jet frame, $\epsilon'$, and the jet-frame flux, $L'_{\epsilon'}$ (in the notation of Z14) need to be multiplied by $(m_{\rm p}/m_{\rm e})^{-2}$, $(m_{\rm p}/m_{\rm e})^{2}$, $(m_{\rm p}/m_{\rm e})^{-1}$ and $m_{\rm p}/m_{\rm e}$, respectively. Correspondingly, the constant $a_{\rm e}$ of equation (33) of Z14 needs to be multiplied by $(m_{\rm p}/m_{\rm e})^{3/2}$, and the power in relativistic particles, $P_{\rm e}$, and in the rest mass, $P_{\rm i}$, need to be multiplied by $(m_{\rm p}/m_{\rm e})^{5/2}$ and $m_{\rm p}/m_{\rm e}$, respectively. The minimum total jet power and the corresponding magnetic field strength of equation (36) in Z14 need to be multiplied by $(m_{\rm p}/m_{\rm e})^{10/7}$ and $(m_{\rm p}/m_{\rm e})^{5/7}$, respectively. 

The method of Z14 assumes a local synchrotron spectrum which can be well represented by a power law in some range of frequencies, and it uses one monochromatic measurement of the flux. Here, we use an estimate of the flux around the maximum of the $\nu F_\nu$ spectrum, using spectral plots in fig.\ 9 of B13. We list our adopted values in Table 1. This implies that the parameter $e_{\rm max}$ of Z14 is $=1$. However, we modify the method in order to take into account that $\gamma_{\rm min}$ is specified in the model of B13. In the case of ion acceleration, one would naturally expect $\gamma_{\rm min}\sim 1$, unlike the case of electrons, which can be preheated to $\langle\gamma_{\rm min}\rangle\gg 1$ (with acceleration only proceeding out of this distribution). However, B13 used $\gamma_{\rm min}=10^3$ for some objects, see Table 1. To account for this, we define $e_{\rm min}=B\gamma_{\rm min}^2/(B_{\rm cr} m_{\rm p}^2 \epsilon')$ (cf.\ Z14). Since $B$ is itself calculated by the minimization, this requires a simple iterative calculation.

\begin{figure}
\centerline{\includegraphics[width=5.5cm]{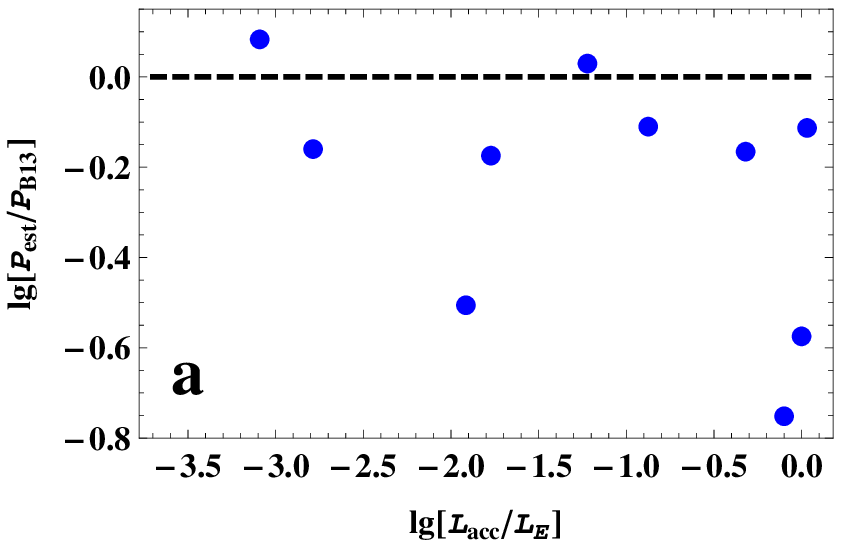}}
\centerline{\includegraphics[width=5.5cm]{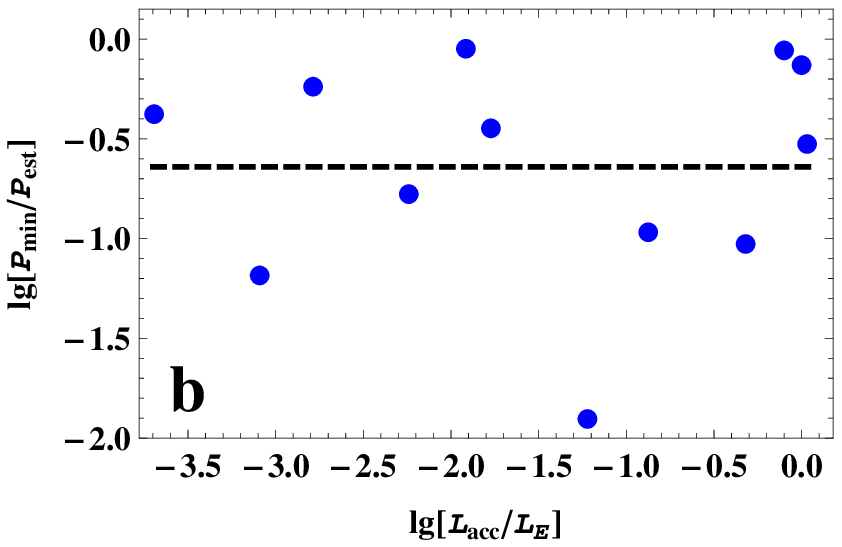}}
\caption{(a) The ratio of the total jet power estimated using the magnetic field strengths of B13 and the method of Z14 to the corresponding power given in B13, as a function of the Eddington ratio. The dashed line corresponds to equal powers. (b) The ratio of the minimum power to that calculated using the method of Z14 with the magnetic field strengths of B13. The dashed line corresponds to the geometric average.
}
\label{ratiob13}
\end{figure}

We first check how well the above method reproduces the jet powers given in B13. For this, we use the values of $B$ given in B13 instead of those corresponding to the minimum power. We show the results in Fig.\ \ref{ratiob13}(a). Some disagreement is expected given that in the present estimate we take into account only the proton synchrotron emission and neglect cooling. The adiabatic cooling is, in fact, dominant in AO 0235+164 and W Comae, in which case the proton-synchrotron estimate gives values $\sim$200 and 50 times too low, respectively. Apart from this, the overall agreement is good, within a factor of 2 for 7 for the remaining objects, see Fig.\ \ref{ratiob13}(a).

We then apply the minimization method to the sample of B13. We have found the minimum powers are on average lower by a factor of few compared to the corresponding estimates above, with the geometric average being $\simeq 0.23$, see Fig.\ \ref{ratiob13}(b). This reduction is due to the minimum corresponding to an equipartition between the power in the relativistic ions, $P_{\rm i}$, and in the magnetic field, $P_B$, which may not be able to produce a fit to the entire SED. Indeed, B13 found $P_{\rm i}\gg P_B$ in most of their fits. This departure from equipartition naturally results in higher total jet powers. Still, the overall possible reduction of the jet power due to changing the fit parameters is by at most a factor of a few.

\section{Astrophysical implications}
\label{implications}

In Section \ref{sample}, we found that the jet powers of the AGN sample of B13 exceed their radiative luminosities by very large factors, with $P_{\rm j}/L_{\rm acc}\sim 50$--$10^5$ and $\langle P_{\rm j}/L_{\rm acc}\rangle\sim 3000$. Then, we showed in Section \ref{minimum} that the values of the jet power obtained by B13 in the hadronic model cannot be significantly reduced by changing the fit parameters, as they are, on average, within a factor of a few of the minimum jet powers estimated using the observed \g-ray spectra for the radiative mechanism dominant in this model, namely proton synchrotron. These findings imply that the jet formation efficiency greatly exceeds that found in leptonic models, which is already quite high \citep[see, e.g.,][]{ghisellini14}. We have also found that the most efficient jet formation mechanism known, based on the Blandford-Znajek mechanism extracting black-hole spin with a magnetically arrested disc \citep{tnm11,mtb12}, is ruled out by the magnetic fields measured at the pc distance scale. Since the jet power has to be then derived entirely from accretion, the implied accretion rates are huge and the radiative efficiencies of the accretion disc must be tiny. 

At the inferred accretion rates, $\langle \dot M c^2/L_{\rm E}\rangle \sim 10^3$, accretion is supercritical (ruling out optically-thin radiatively inefficient models, which can correspond only to $\dot M c^2/L_{\rm E}\ll 1$, see, e.g., \citealt{yn14}). The angle-averaged accretion-disc luminosity in this case is small, corresponding to $L \sim L_{\rm E}$ and a radiative efficiency is $\sim 10^{-3}$ \citep{sikora81,sadowski14}. However, most of that luminosity emerges through axial funnels, where the observed flux is super-Eddington, corresponding to $L \sim 10 L_{\rm E}$ \citep{sikora81,sadowski14}. Our sample consists of blazars seen close to on-axis, on average at $i \sim 1/\Gamma_{\rm j}$, and the accretion-disc-emission funnels have opening angles greater than that \citep{sadowski14}. Thus, we would expect to see super-Eddington accretion-disc luminosities, while, in fact, they are all $\la L_{\rm E}$, with 8 out of 12 objects having $L_{\rm acc} \la 0.1 L_{\rm E}$. Thus, the hadronic emission model for the jets is inconsistent with the standard accretion theory.

The inferred extreme accretion rates present also a major problem in light of results of studies of supermassive-black-hole growth. The e-folding time by which the black-mass would increase due to accretion, $M/\dot M$, is $\simeq 4\times 10^5$ yr at $\langle \dot M c^2/L_{\rm E}\rangle\simeq 10^3$ found in Section \ref{sample}. This is a very short time compared to estimated lifetimes of active phases of radio sources, e.g., $\simeq 2\times 10^8$ yr for FR IIs (\citealt*{antognini12} and references therein). Correspondingly, the average radiative accretion efficiency we found (under the assumptions of the hadronic model) of $\langle L_{\rm acc}/\dot M c^2\rangle \simeq 4\times 10^{-5}$ is $\ll$ than the average accretion efficiency of $\sim$0.1--0.3 \citep{soltan82,marconi04,silverman08,schulze15}. Thus, blazars radiating via the hadronic emission mechanism have to represent at most a small fraction of accreting supermassive black holes and/or such extreme accretion episodes must be extremely short-lived, representing only a duty-cycle of order $\sim 10^{-4}$ (which is in conflict with the fact that the studied objects have quite average properties). 

Finally, the obtained extreme powers are in conflict with studies of the jet power based on radio lobes and X-ray cavities \citep{mh07,cavagnolo10,nemmen12,russell13,gs13,sg13}. Those studies indicate jet powers at most moderately exceeding the Eddington luminosity, and even in the most luminous sources they exceed the accretion-disc luminosity only by a factor of $\la 10$. On the other hand, those powers are in an overall agreement with the powers estimated in leptonic models, see, e.g., \citet{ghisellini14}.

Summarizing, we have found major difficulties in reconciling the jet power requirements of hadronic blazar models with (a) observed accretion-disc luminosities, (b) accretion rates inferred from supermassive-black-hole growth, and (c) jet powers inferred from radio lobes and X-ray cavities. While the problems (b) and (c) can be, in principle, circumvented, if we apply the hadronic model to a small number of sources and assuming a short duty cycle, the problem (a) requires a change of the accretion paradigm for the very sources fitted by the hadronic model. We conclude that our results represent a strong argument against the applicability of hadronic models to blazars. 

\section{Summary}
\label{summary}

We have presented an analysis of the jet powers required by hadronic model fits to the SEDs of blazars, based both on detailed numerical modelling by B13, and on minimal jet power requirements in a proton-synchrotron interpretation of the $\gamma$-ray emission, adopting the methodology of Z14. Our main results are as follows.

We show that hadronic models of B13 that can account for broad-band spectra of blazars emitting in the GeV range, have jet powers of $P_{\rm j} \sim 10^2 L_{\rm E}$, approximately independent of their accretion Eddington ratio, which spans $L_{\rm acc} / L_{\rm E} \simeq 10^{-3}$--1 for the studied sources. The ratio of the jet power to the radiative luminosity of the accretion disc can be as high as $\sim 10^5$ for the studied sample, and $\langle P_{\rm j}/L_{\rm acc}\rangle\sim 3000$.

Furthermore, we show that the available measurements of the radio core shift for the sample imply low magnetic fluxes threading the black hole, which rules out the Blandford-Znajek mechanism with efficient production of jets for hadronic blazar models.

These results require that the accretion rate necessary to power the modelled jets is very high, compared to the accretion radiative output. The average radiative accretion efficiency is $\langle L_{\rm acc} / \dot M c^2\rangle \simeq 4 \times 10^{-5}$ for the studied sample.

A major problem for the the hadronic model is presented by their required highly super-Eddington jet powers. This, in turn requires highly super-Eddington accretion rates, at which the observed luminosities would be much higher than those actually seen. If the model is correct, the currently prevailing picture of accretion in AGNs needs to be revised. 

The extreme jet powers obtained by B13 are in conflict with the estimates of the jet power by other methods. The obtained accretion mode cannot be dominant during the lifetimes of the sources, as the modelled very high accretion rates would result in much too rapid growth of the central supermassive black holes. Thus, if applicable, this accretion mode can only be present with a very small duty-cycle in the black-hole evolution, and thus can be applied only to a very small fraction of the radio loud sources. 

\section*{ACKNOWLEDGEMENTS}

We thank Marek Sikora for valuable discussions, and Patryk Pjanka for help with the data analysis. This research has been supported in part by the Polish NCN grants 2012/04/M/ST9/00780 and 2013/10/M/ST9/00729. MB acknowledges support by the Department of Science and Technology and the National Research Foundation of the Republic of South Africa through the South African Research Chair Initiative (SARChI).

\label{lastpage}

\end{document}